\documentclass[prl,twocolumn,superscriptaddress, preprintnumbers]{revtex4}

\usepackage{times}
\usepackage{amsmath}
\usepackage{amsthm}
\usepackage{amssymb}
\usepackage{amsbsy}
\usepackage{amsfonts}
\usepackage{graphicx}
\usepackage{bibunits}
\usepackage{resizegather}

\def\GL{\mathrm{GL}}

\makeatletter
\let\O\@undefined
\def\O{\mathrm{O}}

\newcommand{\nc}{\newcommand}
\nc{\CA}{{\cal A}} \nc{\CB}{{\cal B}} \nc{\CC}{{\cal C}}
\nc{\CD}{{\cal D}} \nc{\CE}{{\cal E}} \nc{\CF}{{\cal F}}
\nc{\CG}{{\cal G}} \nc{\CH}{{\cal H}} \nc{\CI}{{\cal I}}
\nc{\CJ}{{\cal J}} \nc{\CK}{{\cal K}} \nc{\CL}{{\cal L}}
\nc{\CM}{{\cal M}} \nc{\CN}{{\cal N}} \nc{\CO}{{\cal O}}
\nc{\CP}{{\cal P}} \nc{\CQ}{{\cal Q}} \nc{\CR}{{\cal R}}
\nc{\CS}{{\cal S}} \nc{\CT}{{\cal T}} \nc{\CU}{{\cal U}}
\nc{\CV}{{\cal V}} \nc{\CW}{{\cal W}} \nc{\CX}{{\cal X}}
\nc{\CY}{{\cal Y}} \nc{\CZ}{{\cal Z}}

\nc{\bA}{\mathbb{A}} \nc{\bB}{\mathbb{B}} \nc{\bC}{\mathbb{C}}
\nc{\bD}{\mathbb{D}} \nc{\bE}{\mathbb{E}} \nc{\bF}{\mathbb{F}}
\nc{\bG}{\mathbb{G}} \nc{\bH}{\mathbb{H}} \nc{\bI}{\mathbb{I}}
\nc{\bJ}{\mathbb{J}} \nc{\bK}{\mathbb{K}} \nc{\bL}{\mathbb{L}}
\nc{\bM}{\mathbb{M}} \nc{\bN}{\mathbb{N}} \nc{\bO}{\mathbb{O}}
\nc{\bP}{\mathbb{P}} \nc{\bQ}{\mathbb{Q}} \nc{\bR}{\mathbb{R}}
\nc{\bS}{\mathbb{S}} \nc{\bT}{\mathbb{T}} \nc{\bU}{\mathbb{U}}
\nc{\bV}{\mathbb{V}} \nc{\bW}{\mathbb{W}} \nc{\bX}{\mathbb{X}}
\nc{\bZ}{\mathbb{Z}}

\begin{document}

%%%%%%%%%%%%

\title{Perfect Metal Phases of One-Dimensional and Anisotropic Higher-Dimensional Systems}

\author{Eugeniu Plamadeala}
\affiliation{Department of Physics, University of California, Santa Barbara,
California 93106, USA}
\author{Michael Mulligan}
\affiliation{Microsoft Research, Station Q, Elings Hall,
University of California, Santa Barbara, California 93106-6105, USA}
\author{Chetan Nayak}
\affiliation{Microsoft Research, Station Q, Elings Hall,
University of California, Santa Barbara, California 93106-6105, USA}
\affiliation{Department of Physics, University of California, Santa Barbara, California 93106, USA}

\begin{abstract}
We show that a 1D quantum wire with $23$ channels of interacting fermions has a perfect metal phase
in which all weak perturbations that could destabilize this phase are irrelevant. 
Consequently, weak disorder does not localize it, a weak periodic potential does not open a gap, and contact with a superconductor also fails to open a gap.
Similar phases occur for $N \geq 24$ channels of fermions, except for $N=25$,
and for $8k$ channels of interacting bosons, with $k\geq 3$. Arrays of perfect metallic wires
form higher-dimensional fermionic or bosonic perfect metals, albeit highly-anisotropic ones.
\end{abstract}

\maketitle
%\begin{bibunit}[apsrev]

\paragraph{Introduction.}

Do stable zero-temperature metallic phases exist in one or two dimensions?
A system of non-interacting fermions will always be localized at $T=0$
in dimensions $D=1,2$ in the presence of generic types of impurities \cite{leeramakrishnan85}
\footnote{With spin-orbit interactions and no magnetic field,
there can be a metallic phase in 2D,
but any time-reversal symmetry-breaking perturbation will lead to localization \cite{hikamilarkinnagaoka80}.}.
Localization can be avoided if the fermions have sufficiently strong attractive interactions,
but then they form a superconductor (SC) rather than a metal \cite{giamarchischulz88, RevModPhys.66.261}.
A system of charged bosons is similarly known to have insulating and superconducting phases \cite{PhysRevB.40.546}.
Although the critical point between insulating and superconducting phases is metallic in both
cases, it is not known in either case whether a stable metallic phase exists. 
Such a metallic phase of fermions would necessarily be a non-Fermi liquid since a Fermi liquid becomes localized \cite{PhysRevLett.79.455, PhysRevB.58.R559}
\footnote{The situation is
a little more complicated for spinful fermions. In the presence of disorder, weak interactions
in the spin-triplet channel grow initially at longer scales, but this is likely to lead to magnetic
order (at least locally), followed by localization as for spinless fermions.}.

In addition, we consider a second, related question: if an
infinite array of one-dimensional ($1D$) Luttinger liquids is coupled,
is there a completely stable, albeit anisotropic, non-Fermi liquid phase?
At the turn of the millennium, it was shown that inter-chain interactions could stabilize ``sliding
Luttinger liquid phases'' against many types of interactions
\cite{kivelsonfradkinemergynature, PhysRevLett.80.4341, PhysRevLett.81.5704, PhysRevLett.80.4345, PhysRevLett.83.2745, PhysRevLett.85.2160,PhysRevLett.86.676,PhysRevB.64.045120}.
On physical grounds, one could argue that any other perturbation would be negligibly small and, therefore,
would not have any effect until extremely low temperatures were reached.
But, as a matter of principle, it is not known whether ``sliding Luttinger liquid phases" are
actually stable against \emph{all} perturbations that might cause them to become
superconducting, insulating, or $2D$ Fermi liquids (in which case,
they would become localized by disorder).
%Could remove everything after "perturbations"
Moreover, these constructions
did not lead to completely stable $1D$ metallic systems with a
finite number of Luttinger liquid channels.
Therefore, as a question of principle, it is not known whether there is a
completely stable zero-temperature $1D$ multi-channel Luttinger liquid phase or an
anisotropic $2D$ phase of coupled Luttinger liquids.

In this paper, we answer both questions in the affirmative. We show that there are
one-dimensional phases of interacting fermions and bosons that are stable against
all weak perturbations. 
These phases do not depend upon a symmetry for their stability, unlike the edges of symmetry-protected topological phases \cite{RevModPhys.82.3045, RevModPhys.83.1057, Chen11a, Chen11b}.
They are stable not only against all types of disorder, but also against
coupling to an external 3D superconductor. 
%These phases are stable
%even in the absence of charge conservation, unlike the edges of symmetry-protected
%topological phases \cite{RevModPhys.82.3045, RevModPhys.83.1057, PhysRevB.86.115131}, which %depend upon a symmetry for their stability. 
Since long-ranged order is impossible in 1D \cite{PhysRev.158.383, PhysRevLett.17.1133, Coleman:1973ci}, the absence of proximity-induced
superconductivity is a reasonable definition of `non-superconducting'.
Due to its extreme stability, we call such a phase a \emph{perfect metal}.
If we form an array of perfect metal wires, such an array is a
highly-anisotropic 2D non-Fermi liquid metal or Bose metal \cite{PhysRevB.75.235116}.

These results are based on a relation that we demonstrate
between special values of the interaction parameters of a 1D system with
$N$ channels of fermions (or bosons) and $N$-dimensional odd (or even)
unimodular lattices. Vectors in such a lattice correspond to the different possible
chiral excitations of the system, and the square of the length of a vector is
twice the scaling dimension of the operator that creates the corresponding excitation.
A non-chiral excitation is made of excitations of both chiralities;
at special values of the interaction parameters, its scaling dimension is
the sum of the scaling dimensions of the two chiral operators.
Small changes in the interactions away
from these special values mix the two chiralities, thereby
causing small changes in the scaling dimensions.
Systems that correspond to so-called non-root unimodular lattices with no short vectors
%, i.e., non-root unimodular lattices, 
are stable to all weak perturbations because all such interactions are irrelevant in the renormalization group sense. 
The lowest dimension in which such
an odd lattice exists is $N=23$ (the shorter Leech lattice); for even lattices it is $N=24$
(the Leech lattice) \cite{Conway88}.

\paragraph{Setup.}

The stable metallic phases that we describe in this paper are constructed from one-dimensional electronic systems in which the current-current and density-density interactions have been chosen in a particularly novel way.
Such phases can be accessed by perturbing the conventional action describing N channels of free fermions in 1D:
\begin{align*}
S_0 = \int dt dx \Big[\psi^\dagger_{R,I} i(\partial_t - {v_I}\partial_x) \psi_{R,I} + \psi^\dagger_{L,I} i (\partial_t + {v_I}\partial_x)\psi_{L,I}\Big],
\end{align*}
where the operator $\psi^\dagger_{R,I}$ ($\psi^\dagger_{L,I}$) creates a right-moving (left-moving) fermion excitation about the Fermi point $k_{F, I}$ ($- k_{F, I}$) in channel $I=1, ..., N$.
The velocity of the $I^\text{th}$ channel of fermions is $v_I$
\footnote{We shall always work in the regime where the dispersion of the fermion modes is linear. Quadratic and higher-order corrections may be systematically incorporated into our formalism.}.

The leading quadratic perturbations couple
$\Psi_{IJ}^{SC}=\psi_{R, I} \psi_{L, J}$
to an external $3D$ charge-$2e$ SC at wavevector $k_{F, I} - k_{F, J}$
or the charge-density-wave (CDW) order parameter
$\rho^{2k_F}_{IJ}=\psi^\dagger_{R, I} \psi_{L, J}$ to a periodic electric
potential at wavevector $k_{F, I} + k_{F, J}$.
Both perturbations are relevant at the free fermion fixed point
and generically lead to a gapped ground state that explicitly breaks
translation invariance and/or charge conservation.

The leading fermion-fermion interactions are density-density and current-current
interactions, parametrized by the symmetric matrix $U_{I, J}$, with $I,J=1,\ldots,2N$:
\begin{multline}
\label{eqn:interactions}
S_\text{int} =  \int dt dx \Big[
U_{I, J} \psi^\dagger_{R, I} \psi_{R, I} \psi^\dagger_{R, J} \psi_{R, J}\\
+ U_{I+N, J+N} \psi^\dagger_{L, I} \psi_{L, I} \psi^\dagger_{L, J} \psi_{L, J}\\ +
2U_{I, J+N} \psi^\dagger_{R, I} \psi_{R, I} \psi^\dagger_{L, J} \psi_{L, J} \Big],
\end{multline}
where we assume throughout that the interaction is short-ranged.
These quartic interactions are marginal at tree level.
If they are added to the free fermion action,
the scaling dimensions of the quadratic SC and CDW perturbations, and also all
higher-body fermion interaction terms will generally change.
Generally, attractive density-density interactions drive SC
perturbations more relevant, while repulsive interactions favor the CDW instability.
Forward-scattering interactions that couple densities of the same chirality mix the
collective modes and renormalize their velocities.

%Could remove this paragraph
%The stable metallic phases that we describe in this paper are accessed by varying the
%short-ranged density-density and current-current interactions parameterized by $U_{I J}$.
%We will describe how the density-density and current-current interactions can be tuned to
%give a {\it completely} stable metallic phase that we term the {\it perfect metal}.

\paragraph{Luttinger Liquids and Non-Root Unimodular Lattices.}

To study the perfect metal, it is convenient to use the Luttinger liquid formalism,
which enables us to treat Eqn.~(\ref{eqn:interactions}) non-perturbatively.
Thus, we introduce a single, chiral boson $\phi_{I}$ ($\phi_{I+N}$)
for each chiral fermion $\psi_{R, I}$ ($\psi_{L, I}$).
Our $N$-channel fermion system can be
described by the following bosonic effective action:
\begin{equation}
\label{eqn:LL-action}
    S = \frac{1}{4\pi} \int dt dx \Big[ K_{IJ} \partial_t \phi_I \partial_x \phi_J - V_{IJ} \partial_x \phi_I \partial_x \phi_J \Big].
\end{equation}
Here, $I=1,\ldots, N$
correspond to right-movers and $I=N+1,\ldots, 2N$ correspond to left-movers.
%Could remove the preceding sentence
$K_{IJ}$ is a $2N\times 2N$ symmetric matrix with integer entries.
Density-density and current-current interactions are parameterized
by the symmetric $2N \times 2N$ matrix $V_{IJ}={v_I}\delta_{IJ} + U_{IJ}$
(with $v_I \equiv v_{I-N}$ for $I>N$).
The Hamiltonian associated with this action is positive semi-definite if and only if
$V_{IJ}$ has non-negative eigenvalues.
In addition, we must supplement the action with a periodicity condition
$\phi_I \sim \phi_I + 2 \pi m_I$, for $m_I \in \mathbb{Z}$.

The free fermion fixed point is described within this formalism by choosing $K = K_{\rm ferm} = \mathbb{I}_N \oplus - \mathbb{I}_N$ and $V_{IJ}={v_I}\delta_{IJ}$, where $\mathbb{I}_N$ is the $N\times N$ identity matrix. 
The operators $\psi^\dagger_{I, R} = \frac{1}{\sqrt{2\pi a}} e^{-i \phi_{I}} \eta_{I}$ and
$\psi^\dagger_{I, L} = \frac{1}{\sqrt{2\pi a}} e^{i\phi_{I + N}} \eta_{I+N}$
create, respectively, right- and left-moving fermions in the $I^{\rm th}$ channel;
$a$ is a short-distance cutoff, and the Klein factor $\eta_I$ satisfies
$\eta_J \eta_K = - \eta_K \eta_J$ for $J\neq K$ in order to ensure anticommutation
relations between fermion operators in different channels. 
%Thus, the SC and CDW interactions take the form,
%\begin{align}
%\delta S_{SC/CDW}^{I, J+N} = & \lambda_{SC}^{I, J+N} \int dt dx\ \cos(\phi^I \pm \phi^{J+N}).
%\end{align} 
The density $j_{I}^0$ and current $j_{I}^1$ in the $I^{\rm th}$ channel are given by
$j_I^\mu = \frac{1}{2\pi}\epsilon^{\mu \nu} \partial_\nu \phi_I$ with $\epsilon^{01} = - \epsilon^{10} = 1$.

A system of hard-core bosons can be re-expressed
in terms of fermions by a Jordan-Wigner transformation.
It can then be bosonized
as above, but with $K_{\rm boson}={\sigma_x}\oplus{\sigma_x}\oplus \ldots \oplus {\sigma_x}$.

It is important to observe that there is still some redundancy in the expression for the Luttinger liquid action.
The field redefinition $\phi_I = W_{IJ} \tilde{\phi}_J$
preserves the periodicity conditions of the fields so long as $W\in GL(2N,\mathbb{Z})$ 
\footnote{A $GL(N, \mathbb{Z})$ transformation is represented by a matrix with integer entries whose determinant has unit modulus.}.
However, this redefinition transforms the action in Eqn. (\ref{eqn:LL-action}) into an action of the same form, but
with $\tilde{K} = W^T K W$ and $\tilde{V} = W^T V W$. 

This seemingly innocuous observation has a surprising consequence.
Consider the operator $\cos(m_I \phi_I)$.
It is a local operator that can be added to the Hamiltonian
if $\frac{1}{2}m_I (K^{-1})_{IJ} m_J$ is an integer.
It could, potentially, open a gap
if its right and left scaling dimensions are equal, i.e., if $m_I (K^{-1})_{IJ} m_J =0$.
%We refer to any vector $m_I$ satisfying this condition as a spin-0 or null vector.
This operator is an irrelevant perturbation if its scaling dimension is greater than two, in which case, 
it will not open a gap in the infrared at weak coupling.
%if its coefficient is sufficiently small.
The operator has scaling dimension
$\frac{1}{2}\sum_{I = 1}^{2N} (m_I)^2$ when $V={v_I}\delta_{IJ}$.
Suppose, instead, that $\tilde{K} = W^T K W$ and $\tilde{V} = W^T V W$
are block-diagonal:
\begin{equation}
\label{eqn:preferred-basis}
\tilde{K} = \begin{pmatrix}{\tilde{K}_R} & 0\\ 0 & -{\tilde{K}_L} \end{pmatrix}\, , \,\,
\tilde{V} = \begin{pmatrix} \tilde{V}_R & 0\\ 0 & \tilde{V}_L \end{pmatrix},
\end{equation}
with positive-definite $\tilde{K}_{R,L}$ and $\tilde{V}_{R,L}$.
Then, the field redefinition $\phi_I = W_{IJ} \tilde{\phi}_J$ allows us to
compute the scaling dimension of $\cos(m_I \phi_I)=\cos(m_I W_{IJ} \tilde{\phi}_J)$:
$(\Delta^R_m, \Delta^L_m) = (\mbox{$\frac{1}{2}$} {\tilde{m}^R}\tilde{K}_R^{-1} \tilde{m}^R, \,
\mbox{$\frac{1}{2}$} {\tilde{m}^L} \tilde{K}_L^{-1} \tilde{m}^L)$,
where $\tilde{m}_J = \tilde{m}^{R (L)}_J = m_I W_{IJ}$ for $J = 1, ..., N$ ($J = N+1, ..., 2N$).
If the off-diagonal blocks in $V$ are non-zero, then the total scaling
dimension $\Delta^R_m + \Delta^L_m$ will generally change, but the
spin $\Delta^R_m - \Delta^L_m$ will remain the same.
(These manipulations are a particular manifestation of the observation
that density-density and current-current interactions can modify the scaling dimensions of operators.)

To understand how a perfect metal phase could exist, in which all such operators are irrelevant,
it is useful to express the above ideas more geometrically.
As described in the Supplementary Online Material,
we can associate the $N$-dimensional integral lattices $\tilde{\Gamma}_{R,L}$, with
positive-definite inner products, to the matrices $\tilde{K}_{R,L}$.
The $K$-matrices are the Gram matrices of the lattices
and basis changes in the lattice
transform the $K$-matrices according to
$K \rightarrow \tilde{K} = W^T K W$.
Since $1 = \text{det}(K_\text{ferm/boson})=\text{det}(\tilde{K})=\text{det}({\tilde{K}_R})\text{det}(\tilde{K}_L)$,
we conclude that $|\text{det}({\tilde{K}_{R,L}})|=1$ 
\footnote{We use the fact that $|\text{det}(\tilde{K}_{R,L})| \geq 1$.}.
Therefore, $\tilde{\Gamma}_{R,L}$ are unimodular lattices.
Consequently, the full matrix $\tilde{K} = {\tilde{K}_R}\oplus-{\tilde{K}_L}$ is associated with
the unimodular lattice $\tilde{\Gamma}_R\oplus \tilde{\Gamma}_L$ of signature $(N,N)$ 
\footnote{An $(M+N) \times (M+N)$ matrix has signature $(M,N)$ if it has $M$ positive eigenvalues and $N$ negative eigenvalues.}. 
%Since $\Gamma_{R,L}$ are unimodular, they are equal to their dual lattices,
%and $\tilde{K}_{R,L}$ are $GL(N,\mathbb{Z})$-equivalent to $\tilde{K}^{-1}_{R,L}$.
%Thus, we can equally-well take $\tilde{K}_{R,L}$ to be the Gram matrices of $\Gamma_{R,L}$.
An operator $\cos(\tilde{m}_J {\tilde{\phi}_J})$ can be associated with
a vector $({{\bf \tilde{v}}_R}, {{\bf \tilde{v}}_L})\in \tilde{\Gamma}_R\oplus \tilde{\Gamma}_L$,
where ${{\bf \tilde{v}}_{R,L}}= \tilde{m}^{R,L}_I {\bf \tilde{f}}_{R,L}^I$
and ${\bf \tilde{f}}_{R,L}^I$ are bases for $\tilde{\Gamma}_{R,L}$ satisfying ${\bf \tilde{f}}_{R, L}^I \cdot {\bf \tilde{f}}_{R, L}^J = (\tilde{f}_{R,L}^I)_a (\tilde{f}_{R,L}^J)_a = (\tilde{K}^{-1}_{R,L})^{I J}$ with $a = 1, ..., N$.
The scaling dimension of the operator is $(\Delta^R_m, \Delta^L_m) =
\left(\mbox{$\frac{1}{2}$}{|{{\bf \tilde{v}}_R}|^2} , \,
\mbox{$\frac{1}{2}$} {|{{\bf \tilde{v}}_L}|^2}\right)$
for block-diagonal $\tilde{V}$, as in Eqn.~(\ref{eqn:preferred-basis}).

Therefore, if there are no ${\bf \tilde{v}}_{R,L}\in \tilde{\Gamma}_{R,L}$ such that
$|{\bf \tilde{v}}_R|^2 = |{\bf \tilde{v}}_L|^2$ and
${1 \over 2}|{\bf \tilde{v}}_{R}|^2 + {1 \over 2}|{\bf \tilde{v}}_{L}|^2 \leq 2$, or, simply,
$|{\bf \tilde{v}}_R|^2 = |{\bf \tilde{v}}_L|^2 \leq 2$, then there
are no relevant or marginal spin-0 perturbations of the Luttinger liquid action
Eqn. (\ref{eqn:LL-action}) with the choice of couplings in Eqn.~(\ref{eqn:preferred-basis}).
If, moreover, there are no such ${\bf v}_{R,L}$,
even if $|{\bf \tilde{v}}_R|^2 \neq |{\bf \tilde{v}}_L|^2$, then there are no marginal or relevant
perturbations of any kind
\footnote{The absence of ${1 \over 2}|{\bf v}_{1}|^2 + {1 \over 2}|{\bf v}_{2}|^2 \leq 2$
is the condition for spatially-uniform
perturbations. For quenched random perturbations, the condition is less
stringent: we merely need the absence of vectors
${1 \over 2} |{\bf v}_{1}|^2 + {1 \over 2} |{\bf v}_{2}|^2\leq 3/2$.
For perturbations that act at a single point (that would generate flow to a new boundary condition), the condition is less stringent still:
there must be no vectors ${1 \over 2}|{\bf v}_{1}|^2 + {1 \over 2}|{\bf v}_{2}|^2 \leq 1$.}.
A lattice $\Gamma$ is called a non-root lattice if all ${\bf v} \in \Gamma$ satisfy $|{\bf v}|^2 > 2$ (a vector with $|{\bf v}|^2 = 2$ is called a root vector).
%(a vector with $|{\bf v_{1}}|^2 + |{\bf v_{2}}|^2 = 1$ or 2 would be a root vector). 
Therefore, we have reduced the problem of finding a metallic state that is stable against
all spin-0 perturbations to the problem of finding a non-root unimodular lattice $\tilde{\Gamma}_R$
whose Gram matrix $\tilde{K}_R$ is related to 
$K=K_{\rm ferm}$ (for a system composed out of fermions)
or $K=K_{\rm boson}$ (for a system composed out of bosons)
according to $\tilde{K}_R \oplus - \tilde{K}_L = W^T K W$ for some
$W\in GL(2N,\mathbb{Z})$ and unimodular $\tilde{K}_L$. This also guarantees the irrelevance
of almost all local chiral perturbations, with some exceptions that
we discuss further below (even though such perturbations cannot open
a gap).

At this point, we make use of two fortuitous mathematical facts. 
The first is that there is a unique signature $(N,N)$ unimodular lattice of each parity,
up to $SO(N,N)$ rotations acting on the basis vectors
\footnote{An $SO(M, N)$ transformation can be represented by an $(M+N) \times (M+N)$ matrix $S_{ab}$ of unit determinant satisfying $S_{b a} \eta_{b c} S_{c d} = \eta_{a d}$ where $\eta = \mathbb{I}_M \oplus - \mathbb{I}_N$.}, where  
a lattice is said to have even parity if the norm-squared of all vectors is even and said
to have odd parity otherwise \cite{Conway88}.
Therefore, any difference between the Gram matrices of two such lattices
can only be due to a difference in choice of basis.
Consequently, all signature $(N,N)$
unimodular $K$-matrices of the same parity
are $GL(2N,\mathbb{Z})$-equivalent
\footnote{The relation between $GL(N,\mathbb{Z})$ and $SO(N-1,1)$ transformations is discussed more fully in \cite{PhysRevB.88.045131, PhysRevB.89.115116} and can be applied immediately to the $GL(2N, \mathbb{Z})$ and $SO(N,N)$ transformations needed for this paper.}.
%Since we are only considering non-chiral wires, we take $M = N$.
%The $SO(N,N)$ rotation $S_{ab}$ relates the bases $(f_a)^I, (\tilde{f}_b)^J$ of two signature $(N,N)$ lattices:
%\begin{align}
%S_{a b} (f_b)^I = W^I_J (\tilde{f}_a)^J,
%\end{align}
%with $W \in GL(2N, \mathbb{Z})$.
In particular, there exists a $W\in GL(2N,\mathbb{Z})$
such that $W^T K_{\rm fermion/boson} W = \tilde{K}_R \oplus - \tilde{K}_L$ for any
positive-definite odd/even unimodular lattice $\tilde{\Gamma}_R \oplus \tilde{\Gamma}_L$ with
Gram matrix $\tilde{K}_R \oplus - \tilde{K}_L$.
The second fact is that there exist positive-definite unimodular lattices that contain no roots.
In fact, for any integer $n$, there exists an $N$-dimensional
positive-definite unimodular lattice
whose shortest vector ${|{{\bf v}}|^2} = n$ \cite{Milnor73}. 
The minimal possible dimension $N$
increases with $n$. 
For $n=3$, the minimal $N=23$ (the shorter Leech lattice), while for
$n=4$, the minimal $N=24$  (the even Leech lattice).
The Gram matrices $K_{\rm sL}$ and $K_{\rm L}$ of these two lattices
are given in the Supplementary Material. 

To summarize, there is a unique signature $(N,N)$ unimodular lattice, up to SO(N,N) transformations.
All associated signature $(N,N)$ unimodular $K$-matrices give the same operator spectrum of conformal spins since these are SO(N,N) invariants.
However, each unimodular $K$-matrix gives a different spectrum of scaling dimensions
because these are not SO(N,N) invariants. 
Non-root unimodular lattices are associated with
theories with no relevant cosine operators.
% To summarize,all signature
%$(N,N)$ unimodular $K$-matrices give the same spectrum of conformal spins since these are %SO(N,N) invariants
%and there is a unique signature $(N,N)$ unimodular lattice, up to SO(N,N) transformations.
%However, each unimodular $K$-matrix gives a different spectrum of scaling dimensions,
%since these are not SO(N,N) invariants. 
%Non-root unimodular lattices are associated with
%theories with no relevant cosine operators.

\paragraph{Shorter Leech Liquid.}
%We now illustrate these ideas.
We first consider the case in which $\tilde{K}_R = \tilde{K}_L = K_{sL}$, which we call the
\emph{symmetric shorter Leech liquid}.
%or, sometimes, the \emph{symmetric shorter Leech liquid}.
We will call block diagonal $\tilde{V}$, shown in Eqn.~(\ref{eqn:preferred-basis}), the
\emph{decoupled surface}. On the decoupled surface,
the minimum scaling dimension of an operator is $3/2$ if it is completely chiral and $3$ if it is spin-0.
Small changes in $\tilde{V}$ can only change these scaling dimensions slightly, so there is
a finite region of parameter space in which all potential gap-generating perturbations
are irrelevant. For block diagonal $\tilde{V}$, we can compute the scaling dimensions
of various perturbations using the $GL(46,\mathbb{Z})$ transformation $W_s$,
given explicitly in the Supplementary Online Material,
that satisfies $W_s^T K_{\rm ferm} W_s = K_{sL} \oplus - K_{sL}$.
Note that there are many possible $GL(46,\mathbb{Z})$ transformations
satisfying  $W_s^T K_{\rm ferm} W_s = K_{sL} \oplus - K_{sL}$
and, therefore, many different possible matrices $V$ that lead to the same
block diagonal $\tilde{V}$. The $W_s$ that we construct in the Supplementary Online Material
is not symmetrical between right- and left-movers, which means that our choice of velocities and
interactions is not parity-invariant. Although this facilitated our calculations, it is not essential
for any of our conclusions. 

Table \ref{table:scaling-dims} lists the scaling dimensions of
the electron creation operators $\psi^\dagger_{R/L, I}$; inter-channel exchange operators $J_{R/L;I,J}^\perp = \psi^\dagger_{R/L, I} \psi_{R/L,J}$; SC and CDW order parameters 
$\rho^{2k_F}_{I J}$ and $\Psi^{SC}_{IJ}$; and quartic inter-channel interactions in the
particle-hole channel, ${\cal O}^{\text{p.-h.}}_{IJ} \equiv \psi^\dagger_{L, I} \psi_{R, I} \psi^\dagger_{R, J} \psi_{L, J} $,
and particle-particle channel, ${\cal O}^\text{p.-p.}_{IJ} \equiv \psi_{R, I} \psi_{L, I} \psi^\dagger_{L, J} \psi^\dagger_{R, J}$.
We have indicated the channel indices at which the minimal scaling dimension is obtained for each operator.
Note that the operator $\rho^{2k_F}_{I J}$ scatters a left-moving fermion
in channel $J$ to a right-moving fermion in channel $I$. 
%It is at wavevector
%$k_{F,I} + k_{F,J}$ or, simply, $2k_F$ if the Fermi momentum
%is the same in all channels. 
As noted in the Table 1 caption,
the inter-channel $I=2, J=4$ CDW order parameter has
lower scaling dimension than in any single other channel.
We also see that the most relevant operator is the $2k_F$ charge-density-wave order
parameter in channel $5$.
All of these operators have very high scaling dimensions.
The most relevant operator with $4$ fermion fields
is $\psi_{R,2} \psi^\dagger_{L,2} \psi^\dagger_{R,4} \psi_{L,3}$,
with scaling dimension $10$. Note that operators of this
form destabilize the sliding Luttinger liquid phase in large parts of the phase
diagram \cite{PhysRevLett.86.676}.

The lowest dimension operators are very
complicated combinations of the original electrons.
From the $\theta$-function for the shorter Leech lattice \cite{Conway88}, we can
see that there are 4600 fermionic dimension-$3/2$ operators of each chirality.
One simple (in the tilded basis) dimension-$3/2$ chiral operator
is $e^{i\tilde{\phi}_1}$, but this has a very complicated form in
terms of fermion operators (given in the Supplementary Online Material)
and has total electric charge $-201$.
There are $(4600)^2$ dimension-$3$ operators.
A relatively simple dimension-3 operator (given in the Supplementary Online Material)
is a combination of 10 fermion creation and 12 fermion annihilation operators.

There are also dimension-$(1,0)$ and $(0,1)$
fields $\partial \tilde{\phi}_I$. These shift the Fermi momenta. By coupling such operators
together, we can change the matrix $\tilde{V}_{IJ}$, which is a marginal deformation of
the phase. If such a deformation moves the system off the decoupled surface,
it will change the scaling dimensions of cosine operators, but will leave their conformal
spins unchanged. On the decoupled surface, there are dimension-$(2,0)$ and $(0,2)$
chiral operators -- in fact, 93150 of each \cite{Conway88}. An example is given in the Supplementary Online Material.
They are strictly marginal, due to their chirality, and,
so long as they are sufficiently small, they will not make any of the irrelevant
operators relevant. Hence, they do not destabilize
the shorter Leech liquid, but their coefficients can be non-zero and
they can play a role in determining physical properties on the decoupled surface.
Off the decoupled surface, such an operator will have scaling dimension
$(2+\alpha,\alpha)$ or $(\alpha,2+\alpha)$ and will, therefore,
be irrelevant. These observations also apply to the other perfect metals described in this paper.

\begin{table}[tb]
\begin{tabular}{c|c|c|c|c|c|c|c|c}
 &
$\Delta^{\psi}_{R,I}$ & $\Delta^{\psi}_{L,I}$ &
$\Delta^{J_\perp}_{R,IJ}$ & $\Delta^{J_\perp}_{L,IJ}$ &
$\Delta^{2{k_F}}_{IJ}$ &  
$\Delta^{SC}_{ IJ}$ &
$\Delta^\text{p.-h.}_{IJ}$ &
$\Delta^\text{p.-p.}_{IJ}$ \\
\hline
s& 11/2 & 17/2 & 23 & 13 & 5 & 28 &17 & 21\\
\hline
a& 9/2 & 1/2 & 20 & 1 & 5 & 5 & 113 & 5
\end{tabular}
\caption{The scaling dimensions of various physical operators in the symmetric (s)
and asymmetric (a) shorter Leech liquids. The scaling dimensions depend on the
channel indices $I,J$. We have listed the minimal possible scaling dimensions,
which are attained by $\psi_{R,4}$, $\psi_{L,3}$;  $J^\perp_{R;2, 4}$, $J^\perp_{L;3, 4}$;
$\rho^{2k_F}_{5, 5}$; $\Psi^\text{SC}_{2,4}$;
${\cal O}^\text{p.-h.}_{2,5}$; ${\cal O}^\text{p.-p.}_{3,4}$ in the symmetric case and
$\psi_{R;2}$ $\psi_{L;21}$; $J^\perp_{R;2,5}$, $J^\perp_{L;21,22}$;
$\rho^{2k_F}_{2,21}$; $\Psi^\text{SC}_{2,21}$; 
${\cal O}^\text{p.-h.}_{2,5}$; ${\cal O}^\text{p.-p.}_{2,5}$
in the asymmetric case. The right and left scaling dimensions are not equal even in the symmetric
case, due to the asymmetry in the choice of interactions and velocities, which is not fundamental but was for calculational convenience.}
\label{table:scaling-dims}
\end{table}

\paragraph{Asymmetric Shorter Leech Liquid}

We now consider the case in which $\tilde{K}_R = K_{sL}$ but $\tilde{K}_L = \mathbb{I}_{23}$, 
which we call the \emph{asymmetric shorter Leech liquid}. On the decoupled
surface, the minimum scaling dimension of a right-moving
chiral operator is $3/2$, but a left-moving chiral operator can have dimension-$1/2$.
While the minimal dimension of a spin-0 operator is $3$, as in the case of the
symmetric shorter Leech liquid, there are strictly marginal operators of
dimension-$(3/2,1/2)$ on the decoupled surface, but they are irrelevant off the decoupled surface.

On the decoupled
surface, we can compute the scaling dimensions
of various perturbations using the $GL(46,\mathbb{Z})$ transformation $W_a$
that satisfies $W_a^T K_{\rm ferm} W_a = K_{sL} \oplus - \mathbb{I}_{23}$
and is given explicitly in the Supplementary Online Material.
They are given in Table \ref{table:scaling-dims}.
It is unclear whether the asymmetric shorter Leech liquid can be adiabatically
connected to the symmetric one through a sequence of perfect metal Hamiltonians
in which all potentially gap-generating perturbations are irrelevant.

\paragraph{Region of Stability of Perfect Metals}

As we tune the interactions away from the decoupled surface of any
perfect metal phase associated with a non-root unimodular lattice,
some of the irrelevant perturbations will decrease in scaling dimension
and will, eventually, become relevant. The parameter space is too large
for us to fully map out the region of stability of either the symmetric or
asymmetric shorter Leech liquids. However, as a representative
example, consider the one-parameter family of symmetric
theories with $\tilde{V}(\lambda)=v {M_s^T}{O_s^T}(\lambda) O_s(\lambda)M_s$. Here, $v$ is a velocity scale and the $SO(23,23)$ rotation $O_s(\lambda)_{a b} = \exp\Big(\lambda (\delta_{a 1} \delta_{b 24} + \delta_{a 24} \delta_{b 1})\Big)$ where $a, b = 1, ..., 46$.
%is $\begin{pmatrix} \scriptstyle{\cosh\lambda} & 
%\scriptstyle{\sinh\lambda} \\ 
%\scriptstyle{\sinh\lambda} &  \scriptstyle{\cosh\lambda} \end{pmatrix}$
%in the $2\times 2$ block formed by $I,J \in \{1,24\}$ and is the identity matrix otherwise.
The matrix $(M_s)^a_{I} = (f_I^{(sL)})^a = K_{IJ}(f^{J, (sL)})^a$  is the matrix given in the Supplementary Online Materials such that
$\tilde{V}(0)={M_s^T}M_s$ is of the form given in Eqn. (\ref{eqn:preferred-basis})
with $\tilde{V}_R=\tilde{V}_L=K_{sL}$. 
The minimal scaling dimension
of a spin-0 operator is $3 e^{-2\lambda}$, which becomes relevant at $\lambda\approx 0.203$,
where the largest change in an element of $\tilde{V}$ is $1.25 v$.

\paragraph{Discussion.}

Thus far, we have focused on fermionic systems. 
However, the same basic strategy applies to
bosonic ones as well. The bosonic system associated with the Leech lattice, the lowest dimension
non-root even unimodular lattice, is stable against all weak spin-0 perturbations, since their
minimal scaling dimension is $4$. We will call this phase the Leech liquid.
If we consider systems with more channels, then even the minimal dimension
chiral perturbations are irrelevant. In 48 dimensions, there are 4 lattices with minimal norm
$6$. Moreover, in the $n=8k$ channel asymmetric fermionic case, it is possible for the right-moving sector to be associated with
an even lattice so that all right-moving excitations are bosonic.

Perfect metals are described by conformal field theories (CFTs) with no primary operators of low scaling dimension.
CFTs with a large gap in the spectrum of operator scaling dimensions must have large central charge,
according to Hellerman's inequality $0< \Delta_{\rm min} < (c+\overline{c})/12 + 3/2\pi$ \cite{Hellerman:2009bu}. This may explain why our phases
have a large number of channels. According to the AdS$_3$/CFT$_2$ correspondence \cite{Maldacena:1997re},
such CFTs correspond to weakly-curved gravity duals without light Banados-Teitelboim-Zanelli black holes \cite{Banados:1992wn}.

%It remains an open question whether systems with
%fewer channels can also have stable perfect metal phases. Although is it not possible
%in a single-channel system, it may be possible with $1< n \leq 23$ (fermions) or
%$24$ (bosons) channels.

If we couple a Fermi liquid lead to a point in the middle of a symmetric shorter Leech wire then,
on the decoupled surface, the tunneling conductance will be $G_\text{tun}\sim T^{10}$ due to the high scaling dimension
of electron operators; in an asymmetric shorter Leech wire, it will be Ohmic, $G_\text{tun}\sim T^0$, as
in a Fermi liquid, due to the left-moving sector. These exponents vary continuously as we move away from the decoupled surface.
Other properties are proportional to high powers of $T$ due to the high scaling dimensions of
the operators in Table \ref{table:scaling-dims}.

An array of 1D symmetric shorter Leech or Leech liquids forms an
anisotropic 2D perfect metal. 
Since the minimal
scaling dimension of any quasiparticle creation operator in each 1D wire is $3/2$ (fermions, shorter Leech)
or $2$ (bosons, Leech), all couplings between wires are irrelevant except for
the marginal couplings between densities and currents on the different perfect metal wires.
The irrelevance of tunneling operators precludes the possibility of
charge transport between wires, but density-density and current-current
interactions will enable inter-wire energy transport.
Although inter-wire density-density and current-current
interactions can change the dimensions of cosine operators, the latter are highly irrelevant
in the limit of decoupled wires, so there is a non-zero range of parameter space within which couplings between cosine operators remain irrelevant.

An array of asymmetric shorter Leech liquids presents an even more interesting possibility.
The left-moving channels are chiral Fermi liquids at the decoupled point, and interwire
couplings will drive a crossover to a 2D chiral Fermi surface. On the other hand, the right-moving
channels are chiral shorter Leech liquids, and inter-wire tunneling operators are irrelevant.
Such a system could combine 2D Fermi liquid properties with 1D shorter Leech liquid
properties and exhibit interesting non-Fermi liquid behavior.

\acknowledgements
We would like thank J. Cano, M. Freedman, A. Shapere, B. Ware, and M. Watkins for helpful discussions.
C.N. and E.P. were partially supported by AFOSR under grant FA9550-10-1-0524.

\bibliography{perfect-metal}

%\end{bibunit}

% This file was converted to LaTeX by Writer2LaTeX ver. 1.0.2
% see http://writer2latex.sourceforge.net for more info

%\renewcommand*{\citenumfont}[1]{S#1}
%\renewcommand*{\bibnumfmt}[1]{[S#1]

%%%%%%%%%%%%%%%%%%%%%%%%%%%%%%%%%%%%%%%%%%%%%%%%%%%%%%%%%%%%%%%%%%%%%%%%%%%%%
%\renewcommand{\thesection}{S.\arabic{section}}
%\renewcommand{\thesubsection}{\thesection.\arabic{subsection}}
\setcounter{equation}{0}
\setcounter{figure}{0}
% Hack for making SOM Equations Conform to Science Format
%
% e.g. (S1), (S2), etc
% Requires AMS
\makeatletter %% With ams
\def\tagform@#1{\maketag@@@{(S\ignorespaces#1\unskip\@@italiccorr)}}
\makeatother
% Hack for making figures Say \figurename S\thefigure, e.g. Figure S1:
\makeatletter
\makeatletter \renewcommand{\fnum@figure}
{\figurename~S\thefigure}
\makeatother

% use bibnumfmt to change style at the end of the document
\renewcommand{\bibnumfmt}[1]{[S#1]}
% citenumfont command adds S to all numbers
\renewcommand{\citenumfont}[1]{S#1}
 
\renewcommand{\figurename}{Figure}
%\begin{bibunit}[apsrev]
%%%%%%%%%%%%%%%%%%%%%%%%%%%%%%%%%%%%%%%%%%%%%%%%%%%%%%%%%%%%%%%%%%%%%%%%%%%%%

\section{Supplemental Materials}

\subsection{Relation Between $K$-matrices and Lattices}
\label{sec:lattices}

We now make the relation between $K$-matrices and lattices, used in the
main body of the paper, more explicit.
Let $\lambda^{(R,L)}_a$ for $a=1,\ldots, N$
be the eigenvalues of $({K^{-1}_{R,L}})^{IJ}$ with 
corresponding eigenvectors $({f^{(R,L)}_a})^I$.
We normalize the eigenvectors so that 
$(K^{-1}_{R,L})^{IJ} = ({f^{(R,L)}_a})^I ({f^{(R,L)}_a})^J$ (this construction is identical to the introduction of vielbeins in general relativity).
Now suppose that we view the $({f^{(R,L)}_a})^I$ as the components of a
vector ${\bf f}_{R,L}^I \in \mathbb{R}^N$.
To make this more concrete, define
the unit vectors ${\bf \hat{x}_a} = (0, ..., 0, 1, 0, ..., 0)^{{\rm tr}}$ with a $1$ in the a-th entry and zeros otherwise.
They form an orthonormal basis of $\mathbb{R}^N$
so that ${\bf \hat{x}_a}\cdot {\bf \hat{x}_b} \equiv \delta_{ab}$.
Then we can define ${\bf f}_{R,L}^I \equiv ({f^{(R,L)}_a})^I {\bf \hat{x}_a}$.
Thus, the eigenvectors ${\bf f}_{R,L}^I$ define a lattice $\Gamma_{R,L}$ in $\mathbb{R}^N$
according to $\Gamma_{R,L} = \{\tilde{m}_{I} {\bf f}_{R,L}^I  | \tilde{m}_{I} \in \mathbb{Z}\}$.
This lattice has inner product
${\bf f}_{R,L}^I \cdot{\bf f}_{R,L}^J = ({K^{-1}_{R,L}})^{IJ}$, i.e.
$K^{-1}_{R,L}$ is the Gram matrix of this lattice.
Since $\Gamma_{R,L}$ are unimodular, they are equal to their dual lattices,
and $K_{R,L}$ are $GL(N,\mathbb{Z})$-equivalent to $K^{-1}_{R,L}$.
Thus, we can equally-well take $K_{R,L}$ to be the Gram matrices of $\Gamma_{R,L}$.

\subsection{Finding $W$-matrices}
\label{sec:W-matrices}

There is a recursive procedure for finding the $\GL(N,\bZ)$ transformation that transforms any unimodular $K$-matrix of signature $(N,N)$ to $\mathbb{I}_N \oplus -\mathbb{I}_N$ \cite{watkins2013some}. We describe a single iteration of this procedure. Let $K$ be the $K$-matrix.

{\bf Step 1}: Find a vector of integers $\vec{v}$ such that $\vec{v} \cdot K \cdot \vec{v}^T = 0$. 

{\bf Step 2}: Find a vector of integers $\vec{w}$ such that $\vec{v} \cdot K \cdot \vec{w}^T = 1$, and let $k = \vec{w} \cdot K \cdot \vec{w}^T$.

{\bf Step 3}: Choose any set $\vec{e}_i$ of vectors spanning $\bZ^{2N}$. If $n_i = \vec{e}_i \cdot K \cdot \vec{v}^T \neq 0$, then shift $\vec{e}_i \mapsto \vec{e'}_i = \vec{e}_i - n_i \vec{w}$, such that $\vec{e'}_i \cdot K \cdot \vec{v}$ = 0. Similarly, if $m_i = \vec{e'}_i \cdot K \cdot \vec{w}^T \neq 0$, then shift $\vec{e'}_i \mapsto \vec{e''}_i = \vec{e'}_i - m_i \vec{v}$, such that $\vec{e''}_i \cdot K \cdot \vec{w}$ = 0. Note that this is possible exactly because $\vec{v} \cdot K \cdot \vec{v}^T = 0$ and $\vec{v} \cdot K \cdot \vec{w}^T = 1$. 

One then eliminates any two vectors from $\{ \vec{e''}_i \}$ that are not linearly independent from the rest. Call this final set $\{ \vec{u}_i \}$. Construct a $\GL(2N,\bZ)$ transformation out of the row vectors: $W_1 =  \{ \vec{v}, \vec{w}, \vec{u}_1, \vec{u}_2,... \}$. In the new basis, the $K$-matrix now looks block-diagonal, with $K'$ of dimensions $(2N-2) \times (2N-2)$:
\begin{equation*}
    W_1^T K W_1 = \begin{pmatrix} 0 & 1 & 0 \\ 1 & k & 0 \\ 0 & 0 & K'  \end{pmatrix}
\end{equation*}

This procedure can now be repeated on $K'$ to obtain $W_2$ (appropriately enlarged by $\mathbb{I}_2$ to make it of size $2N \times 2N$), and so on $N-1$ times. 

The composite transformation $W = W_1 W_2 ... W_N$ diagonalizes the original $K$ into $N$ unit determinant blocks of dimension $2 \times 2$.

These blocks can be further diagonalized with $\GL(2,\bZ)$ transformations as follows. If the block has an odd on the diagonal, then
\begin{eqnarray*}
M = \begin{pmatrix} 0 & 1 \\ 1 & k \end{pmatrix} &,& U= \begin{pmatrix}
a & b \\ c & d \end{pmatrix} \\
ac-bd=1 &,& c^2-d^2=k \\
U^T M U &=& \begin{pmatrix} 1 & 0 \\ 0 & -1 \end{pmatrix}
\end{eqnarray*}
Otherwise, if $k$ is even:
\begin{eqnarray*}
M = \begin{pmatrix} 0 & 1 \\ 1 & k \end{pmatrix} &,& U= \begin{pmatrix}
a & b \\ c & d \end{pmatrix} \\
ad+bc=1 &,& 2cd=k \\
U^T M U &=& \begin{pmatrix} 0 & 1 \\ 1 & 0 \end{pmatrix}
\end{eqnarray*}

At this point our $K$-matrix is a direct sum of $\sigma_x's$ and $\sigma_z's$. Finally we can bring $K$ into the form $\sigma_z \oplus ... \oplus \sigma_z$ through applying the following $\GL(4,\bZ)$ transformation as needed:
\begin{eqnarray*}
\sigma_z \oplus \sigma_z & =& U^T  \left( \sigma_z \oplus \sigma_x \right)U   \\
U &=& \left(
\begin{array}{cccc}
 1 & 0 & -2 & -2 \\
 0 & 1 & -1 & -1 \\
 2 & -1 & -1 & -2 \\
 0 & 0 & 1 & 1 \\
\end{array}
\right)
\end{eqnarray*}

Explicit forms for the matrices $K_{SL}$, $W_s$, $W_a$, and $K_L$ as well as the vectors that define the operators mentioned in the text are given in the Mathematica file (and also in a text file) in the supplementary material.

%\end{bibunit}

\end{document}